\documentclass[aps,prl,groupedaddress,amsmath,amssymb,notitlepage]{revtex4-1} 
\usepackage{graphicx}  
\usepackage{dcolumn}   
\usepackage{bm}        
\usepackage{verbatim}   
\usepackage{upgreek}

\usepackage[sort&compress]{natbib} 		

\usepackage[colorlinks=true,
			linkcolor=blue,
			citecolor=blue,
			urlcolor=blue]{hyperref}

\begin{document}

\title{Convolutional Neural Networks for Real-Time Localization and Classification in Feedback Digital Microscopy}
\author{Martin Fr\"anzl and Frank Cichos}
\affiliation{Molecular Nanophotonics Group, Peter Debye Institute for Soft Matter Physics, Universit\"at Leipzig, Linnestr. 5, 04103 Leipzig, Germany}

\date{\today}
\begin{abstract}
We present an adapted single-shot convolutional neural network (YOLOv2) for the real-time localization and classification of particles in optical microscopy. As compared to previous works, we focus on the real-time detection capabilities of the system to allow for a manipulation of microscopic objects in large heterogeneous ensembles with the help of feedback control. The network is capable of localizing and classifying several hundreds of microscopic objects even at very low signal-to-noise ratios for images as large as $416 \times 416$ pixels with an inference time of about 10~ms. We demonstrate the real-time detection performance by manipulating active particles propelled by laser-induced self-thermophoresis. In order to make our framework readily available for others we provide all scripts and source code. The network is implemented in Python/Keras using the TensorFlow backend. A C library supporting GPUs is provided for the real-time inference.
\end{abstract}

\maketitle


\section{Introduction}

Optical microscopy provides structural information but also allows to follow dynamical processes from single molecules and single particles to cells and tissues. Images with high spatial, temporal and also spectral resolution may be obtained. Especially the ability to see dynamic processes opens the possibility to influence these processes in real time via feedback processes. In the field of single molecule detection this has been demonstrated with the electrokinetic or the thermophoretic trap \cite{Cohen2006, Braun2015, Fraenzl2019}. In both cases the optical images are analyzed in real time to extract particle or molecule positions to control electric or temperature fields for positioning purposes. Similarly, feedback control is able to explore new physics in optical tweezers or control active particles by specific rules \cite{Qian2013, Bregulla2014, Khadka2018}. The latter field experiences a quickly growing interest \cite{Bechinger2016, Baeuerle2018, Cichos2020} and will require new image analysis techniques which go beyond algorithmic approaches. 
The main requirements for those new approaches are i) to be able to process images at video rate, ii) the ability to differentiate between multiple species, iii) to work at different optical contrast and signal-to-noise ratios. These requirements are often met by algorithmic approaches using thresholding and centroid calculation or even more advanced versions. Yet, the more complex the image is, e.g., having particles with different contrast, the bigger is the computational effort that has to be spend at the cost of speed \cite{Wang2016, Yucel2017, Kapoor2019}.
Recently, machine learning methods have been introduced to the field of optical microscopy and single particle detection. Those methods are used for image segmentation, holographic reconstruction or also particle tracking \cite{Rivenson2017, Newby2018, Hannel2018, Helgadottir2019, Rivenson2019, Pinkard2019}. Methods for particle and object tracking currently employed in digital microscopy are based on deep convolutional neural networks designed for post-processing, i.e., they are optimized for accuracy, not speed. Therefore, the use of these networks in feedback controlled manipulation in optical microscopy is not feasible.

\begin{figure*}[!h]
\centering
\includegraphics[width=\linewidth]{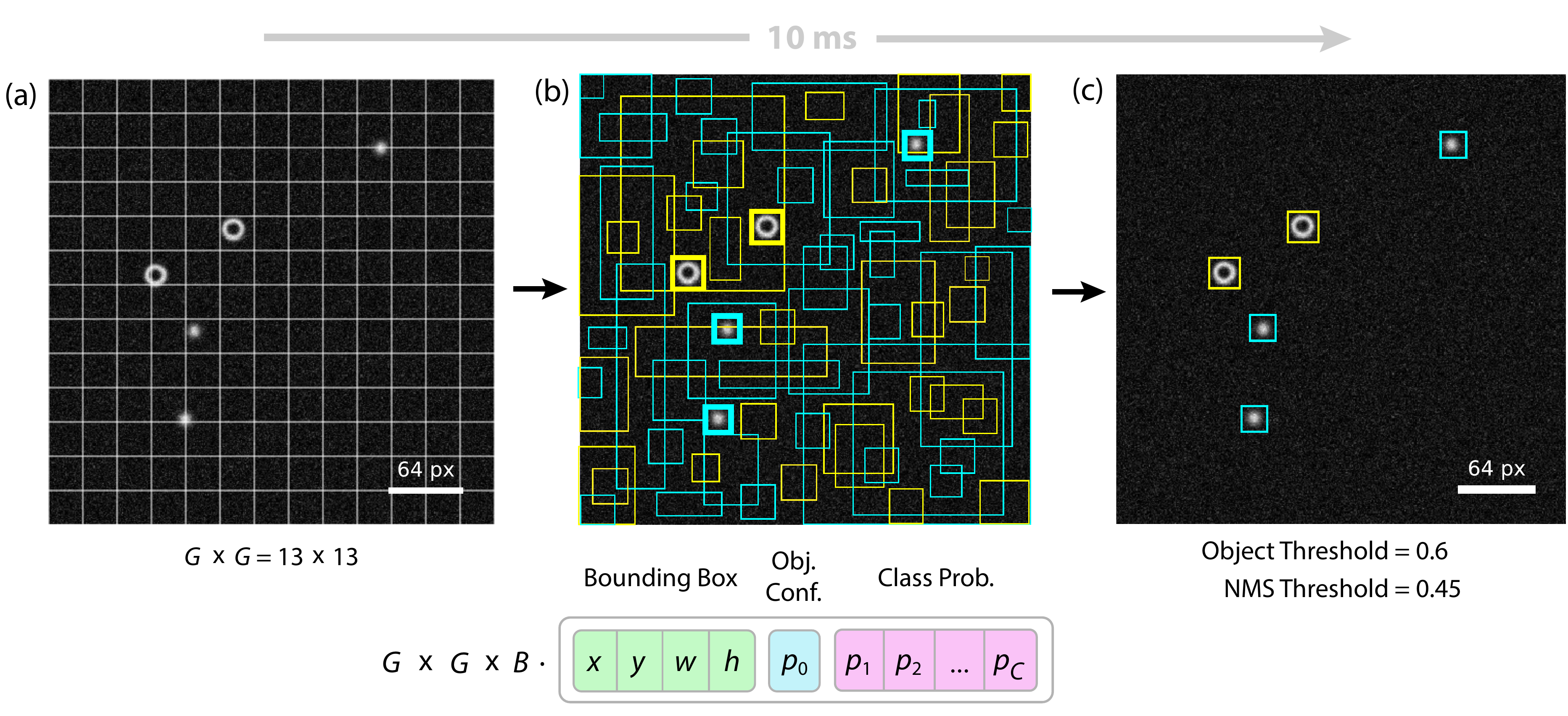}
\caption{Detection principle. (a) The network takes an input RGB image of size 416 $\times$ 416 pixel and divides it into a $G \times G$ grid with $G = 13$. (b) For each grid cell it predicts $B$ bounding boxes, confidence for those boxes and $C$ class probabilities. Here we used $B = 5$ and $C = 2$. These predictions are encoded in a $G \times G \times B \cdot (4 + 1 + C)$ output tensor. The line thickness of the bounding boxes in (b) depicts the object confidence whereas the color of the bounding box is selected according to the highest class probability. (c) Only bounding boxes with an object confidence larger than a certain object threshold are retained. A non-maximum suppression (NMS) algorithm and a NMS threshold value is used to remove overlapping bounding boxes that belong to the same object. Typical values are 0.6 and 0.45 for the object and NMS threshold, respectively.}
\label{Fig:Figure1}
\end{figure*}

Here, we present a single-shot convolutional neural network for real-time detection and classification in digital microscopy. The network is based on the YOLO architecture (``You  Only Look Once'') \cite{Redmon2016a, Redmon2016b, Redmon2018} enabling to detect and classify objects in microscopy images. The single shot-architecture together with a GPU implementation in LabVIEW allows us to perform the particle localization and detection in real time at a speed of 100~fps for 416 $\times$ 416~pixel sized images. The processing speed is not limited by the number of objects available in the image. We evaluate the accuracy of the network as compared to other approaches and its capability of reflecting physical properties of the system. With the help of this approach we demonstrate the feedback control of active particles mixed samples with passive particles in an optical microscopy setup.

\section{Network Structure, Training and Deployment}

\subsection{Network Structure}
The used single-shot-detection approach is based on the TinyYOLOv2 network architecture \cite{Redmon2016b}. It consists of 9 convolutional layers ({Supplement 1}, Section 1A), where the first layers take an input RGB image of the size of 416 $\times$ 416 pixels. The input image is divided into a 13 $\times$ 13 grid where each grid cell is 32 $\times$ 32 pixels (Fig. \ref{Fig:Figure1}(a)). For each grid cell the output predicts 5 bounding boxes where for each bounding box values for the position and size of an object as well as the confidence of detection and a probability for each class are predicted (Fig. \ref{Fig:Figure1}(b)). Only bounding boxes are used for further evaluation if the confidence of object detection is larger than a certain object threshold (Fig. \ref{Fig:Figure1}(c)). A non-maximum suppression (NMS) algorithm and a NMS threshold value is used to remove overlapping bounding boxes that belong to the same object. The object class is assigned according to the maximum value of the predicted class probabilities. A more detailed description of the output decoding can be found in {Supplement~1}, Section 1B.

\subsection{Training}
The network is trained in Python/Keras using the TensorFlow backend \cite{Abadi2016, Chollet2015, Chollet2018} on a GeForce GTX 1660 Ti GPU without any pretrained weights. The corresponding Python scripts for the training of the network and the generation of synthetic training data are supplied with {Code~1} and explained in detail in {Supplement~1}, Sections 2--4. While the synthetic datasets used in this work resemble darkfield microscopy images of nano- and microparticles, Janus-type as well as rod-like and elliptical microparticles, any other training set may be used. Note that all images are assumed to be in focus without changing the contrast of diffraction patterns when defocusing. We train the network with a training set of $25000$ images and a validation set of $5000$ images for 10 epochs and a batch size of 8. The image generation takes about 30~min on a Intel Core i7 9700K 8 $\times$ 3.60~ GHz CPU and the training process about 1~hour on a GeForce GTX 1660 Ti GPU.

\subsection{Deployment}
The trained network graph is exported and deployed to a LabVIEW program developed in the lab which is controlling our microscopy setup. The LabVIEW implementation comprises dynamic link libraries (DLLs) written in C that take an RGB image as input and deliver the decoded output. To get the fastest possible image processing the DLLs are using the GPU supported TensorFlow C API. The details regarding the software can be found in \cite{GitHub2020a} and {Supplement 1}, Section 4. Using a  GeForce GTX 1660 Ti GPU an inference time of about 10~ms is achieved for RGB images with a size of 416 $\times$ 416 pixels. This inference time might be further improved by using a faster GPU or smaller input image sizes.

\section{Results}

\subsection{Evaluation of the Network Performance}
\begin{figure*}[!h]
\centering
\includegraphics[width=\linewidth]{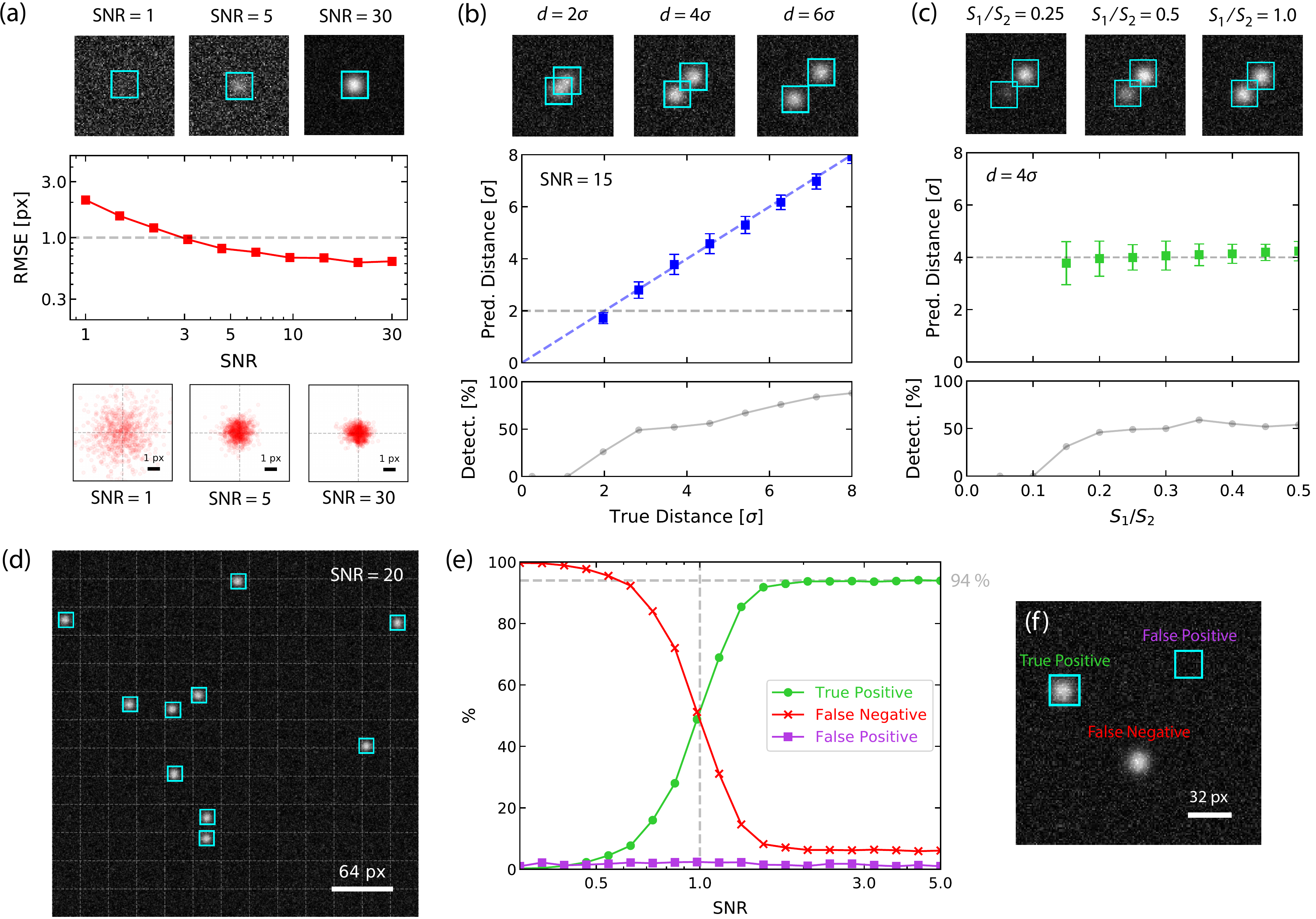}
\caption{Single class training set. (a) Mean-squared error (MSE), i.e., the mean-squared distance between the true and the predicted center position, as function of the signal-to-noise ratio (SNR). Each SNR value was sampled with $S = 1000$ images containing $N = 1$ particle with randomized position. (b) Predicted distance as function of the true distance for SNR = 15. Each distance was sampled with $S = 100$ images containing $N = 2$ particles with randomized position and orientation. The error bars indicate the standard deviation of the predicted distances. The lower graph depicts the percentage of images where two particles have been detected. (c) Predicted distance as function of the intensity ratio averaged over $S = 100$ sample images containing $N = 2$ particles with randomized position and orientation. The error bars indicate the standard deviation of the predicted distances. The lower graph, again, depicts the percentage of images where two particles have been detected. (d) The detection output for an image with $N = 10$ particles and SNR = 20. (e) Percentage of true positives (dots), false negative (crosses) and false positive (squares) detection as function of the SNR. Each SNR value was a sample with $S = 100$ images containing $N = 10$ particles at randomized positions as plotted in (d). (f) Visualization of true positive, false negative and false positive detections.}
\label{Fig:Figure2}
\end{figure*}

The performance of the network is evaluated under different conditions for synthetic datasets of various structure ({Supplement 1}, Sections 3). We evaluate the accuracy of the position detection for single objects, close encounters and the number of false/true positive and negative detections for multiple objects within an image. These parameters are evaluated for datasets with only a single class present and for datasets with multiple classes present. The accuracy is calculated as function of the signal-to-noise ratio (SNR) where the SNR of an image is defined as the ratio of the particles mean signal to the standard deviation of the signal. For each instance the network has been trained with images that contain multiple objects with randomized numbers and SNRs. An overview of the investigated datasets is available in {Supplement 1}, Section 3. If not stated otherwise the network output was decoded with an object threshold of 0.6 and a NMS threshold of 0.45.

\paragraph*{Single Class Training Set}
The first synthetic dataset contains Gaussian spots, which approximate the point spread function of point-like objects in fluorescence or darkfield microscopy (Dataset 1). The width of the spots is constant with $\sigma = 4$~pixels. Details of the dataset as well as a set of sample images are available in {Supplement 1}, Section 3A.
Fig. \ref{Fig:Figure2}(a) depicts the root-mean-square error (RMSE) of the localization as function of the SNR. At each SNR the accuracy was determined from $S = 1000$ test images containing $N= 1$ particle at randomized position. The network has been corrected for a constant offset vector ({Supplement 1}, Section 5). Samples of test images for different SNR values are illustrated in the upper part of Fig. \ref{Fig:Figure2}(a). The lower part of Fig. \ref{Fig:Figure2}(a) depicts point-distribution plots of the error in the position detection for different SNR values. The results show an RMSE for the particle position below one pixel for SNR values larger than 3. For increasing SNR, the error decreases and saturates at a constant value of about 0.6~px. This is contrary to algorithmic approaches, where the RMSE scales with the inverse of the SNR \cite{Chenouard2014}. Thus, algorithmic approaches yield better accuracy for high SNR for single class detection but also a stronger dependence on the SNR. For low SNRs in the range of 1 to 10 our network compares well to the localization accuracy of advanced, algorithmic methods \cite{Chenouard2014}. While recent machine learning approaches have shown ever better performance in terms of the localization accuracy \cite{Helgadottir2019} they are, however, not suitable for real-time approaches.

For real-time processing in microscopy the identification of separate particles in close encounters plays an important role. Fig. \ref{Fig:Figure2}(b) shows the predicted distance of two close particles as function of their true distance for a fixed SNR of 15. Each distance was sampled with $S= 100$ images containing $N = 2$ particles with randomized position and orientation. The error bars indicate the standard deviation of the predicted distances. Samples of test images for different distances $d$ are illustrated  in the upper part of Fig. \ref{Fig:Figure2}(b). The lower graph in Fig. \ref{Fig:Figure2}(b) plots the percentage of test images where two particles have been detected. When detected, the predicted distance nicely reflects the true distance down to a value of 2$\sigma = 8$~pixels. Below that distance, the particles are not recognized as two separate objects. At a distance of $2\sigma$ 25~\% of the sampled particle pairs are identified as two particles. The probability to identify the individual particles increases with the distance and saturates for distances larger than $8\sigma = 32$~pixels at 94~\%. 

Fig. \ref{Fig:Figure2}(c) highlights the performance of the detection when both particles have different contrast as measured by the signal ratio $S_1/S_2$. The particles are kept at a fixed distance of $4\sigma = 16$~pixels with $\mathrm{SNR} = S_2 = 15$. Each $S_1/S_2$ ratio was sampled with $S= 100$ test images containing $N= 2$ particles with randomized position and orientation. The error bars indicate the standard deviation of the predicted distances. Samples of test images for different $S_1/S_2$ ratios are illustrated in the upper part of Fig. \ref{Fig:Figure2}(c). The lower graph in Fig. \ref{Fig:Figure2}(c) plots the percentage of test images where two particles have been detected. When detected, the predicted distance well predicts the true distance ($4\sigma$). The probability to identify the individual particles is increasing with an increasing $S_1/S_2$ ratio and saturates at $S_1/S_2 = 0.2$ at a constant value of about 50~\% (see also Fig. \ref{Fig:Figure2}(b) at $4\sigma$). Thus, even when the second particle is by a factor 6 dimmer than its neighbor, our network detects two particles with the same accuracy as for equal contrast.

Fig. \ref{Fig:Figure2}(d) shows the predicted particle positions for a test image with $N= 10$ particles at a SNR of 20.  In Fig. \ref{Fig:Figure2}(e) the percentage of true positive, false negative and false positive detections (see Fig. \ref{Fig:Figure2}(f) for reference) as function of the SNR are plotted. Each SNR value was a sample with $S= 100$ test images containing $N= 10$ particles at randomized positions as depicted in Fig. \ref{Fig:Figure2}(d) for $\mathrm{SNR} = 20$. The number of true positive detections drops considerably for a signal-to-noise ratio $\mathrm{SNR} < 1$ at the cost of false negatives. At $\mathrm{SNR}= 1$, about 50~\% of the objects are detected, while only about 1~\% are detected false positive. This is remarkable since even at a SNR level of 1 it is difficult to identify objects by eye (see Fig. \ref{Fig:Figure2}(a) for reference). For large SNR the percentage of true positive detections saturates at 94~\% while 6~\% are false negative. For SNR values larger than 2 we observe < 1~\% false positive detections.

As a second example of single class detection we also evaluated the  detection performance for ring-shaped profiles as observed for darkfield images of micrometer-sized particles (Dataset 2). Details of the dataset as well as a set of sample images are available in {Supplement 1}, Section 3B. Fig. S16 shows the evaluation for ring-shaped particles similar to Fig. \ref{Fig:Figure2}. In terms of the localization error (Fig. S16(a)) and the detection for different signal ratios (Fig. S16(b)) the results are comparable to Fig. \ref{Fig:Figure2}(a), \ref{Fig:Figure2}(c). Remarkably, for ring-like particles it is possible to detect overlapping particles (Fig. S16(b)) and 90 \% of the particles are detected true positive even at a SNR of 1 (Fig. S16(e)).

\begin{figure*}[!ht]
\centering
\includegraphics[width=\linewidth]{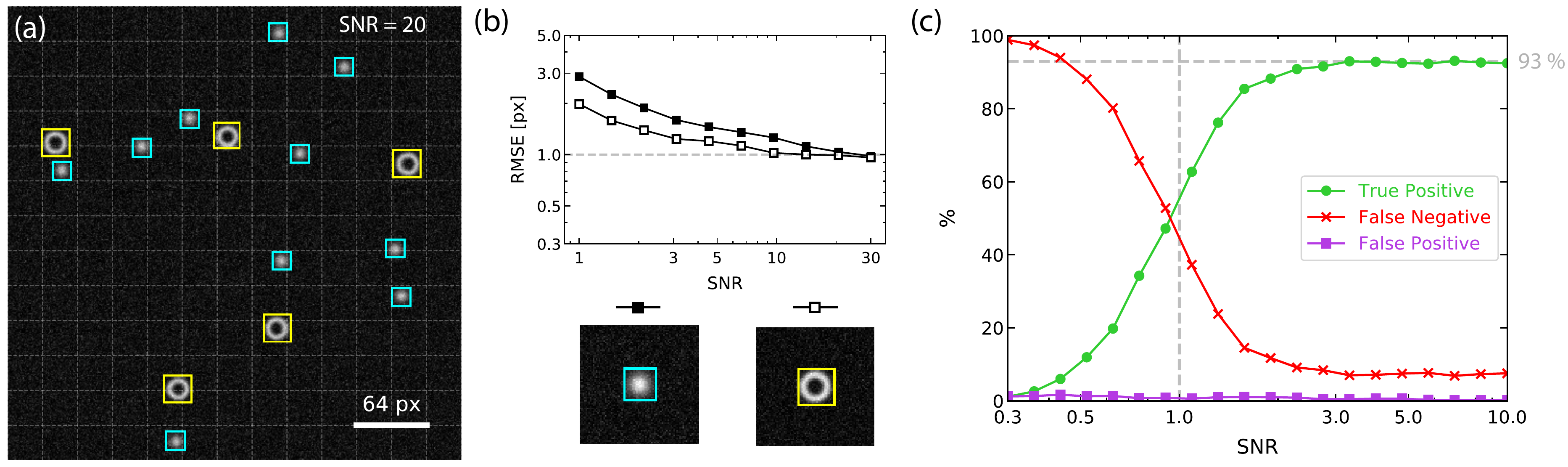}
\caption{Two training classes. (a) The predicted particle locations and classes for a test image with $N_1 = 10$ spots and $N_2 = 5$ ring-shaped particles at a SNR of 20. (b) The mean-squared error of the localization for both particle classes as function of the SNR.  (c) Percentage of true positive (dots), false negative (crosses) and false positive (squares) detections as function of the signal-to-noise ratio. Each SNR value was sampled with $S= 100$ test images containing $N_1= 10$ spots and $N_1= 5$ ring-shaped particles at randomized positions as illustrated in (a).}
\label{Fig:Figure3}
\end{figure*}

\paragraph*{Two Class Training Set}
When training the network with two classes of particle images, e.g., the Gaussian spots and the ring shaped images (Dataset 3, {Supplement 1}, Section 3C), the network slightly looses accuracy as compared to the single class detection. Each SNR value was sampled with $S= 100$ test images containing $N_1= 10$ spots and $N_1= 5$ ring-shaped particles at randomized positions as depicted in Fig. \ref{Fig:Figure3}(a) for $\mathrm{SNR} = 20$.
Fig.  \ref{Fig:Figure3}(a) illustrates an example image with the predicted particle positions $N_1 = 10$ spots and $N_2 = 5$ ring-shaped particles at a SNR of 20. The corresponding localization errors are shown Fig. \ref{Fig:Figure3}(b) as function of the SNR. At a SNR of 3, the RMSE is now slightly above 1~px and a better localization is found for the larger ring shaped objects. With increasing SNR the localization errors saturate at an RMSE of about 1~px. 
Remarkably, neither false positive detections nor classification errors have been observed for the two class dataset.

\begin{figure}[!ht]
\centering
\includegraphics[width=0.42\textwidth]{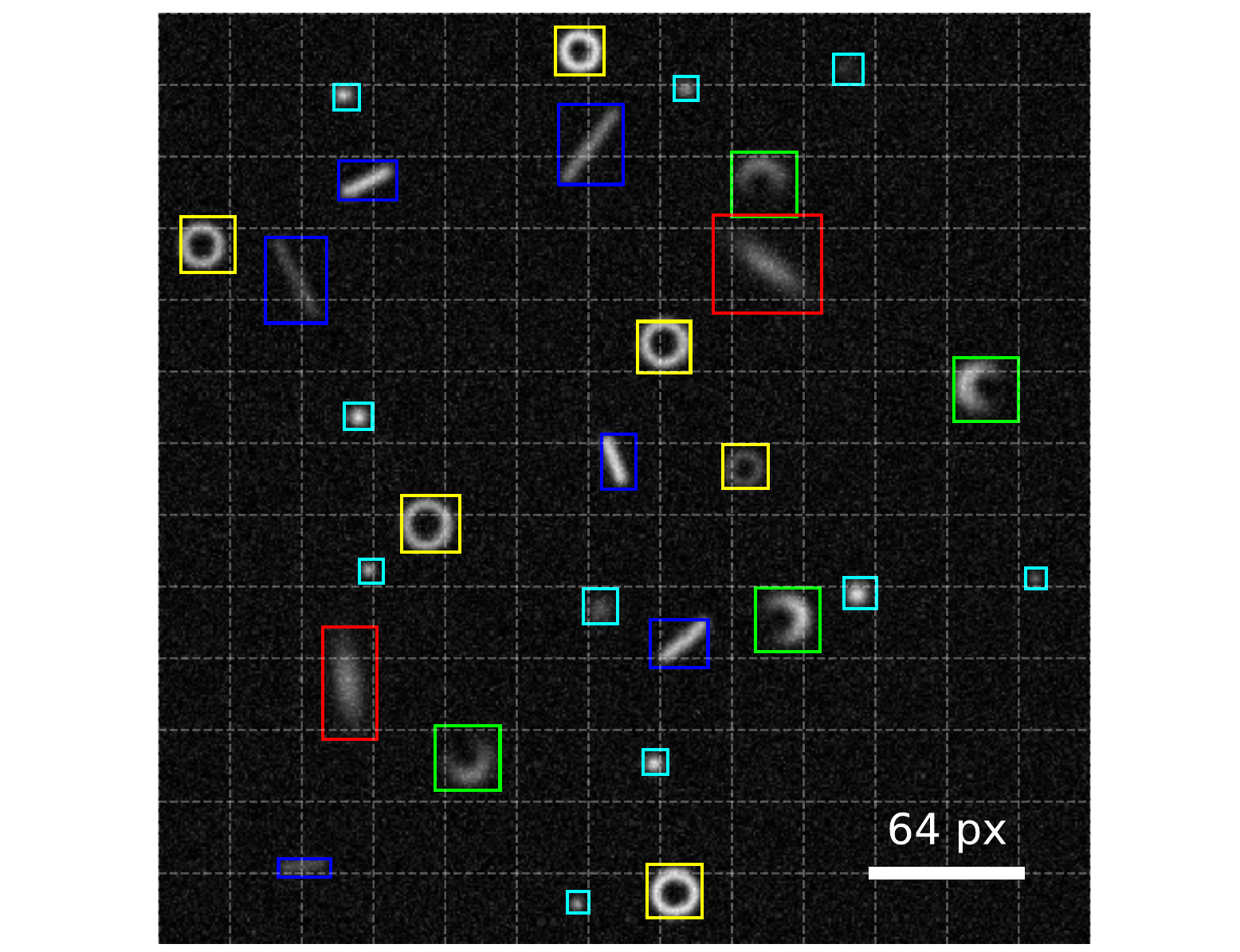}
\caption{Multiple training classes. Predicted locations, sizes/orientations and classes for a test image for a model trained with five different particle classes.}
\label{Fig:Figure4}
\end{figure}

\paragraph*{Multi Class Training Set}
The number of classes to be identified and localized in the image can be extended to more than two classes. The original YOLOv2 model was trained for several thousand object classes \cite{Redmon2016b}. Situations with multiple classes are for conventional algorithmic localization and classification very challenging \cite{Crocker1996, Shen2000, Cheezum2001, Rogers2007, Anthony2008, Parthasarathy2012, Chenouard2014, Wang2016, Yucel2017, Kapoor2019} even if the individual particles have a high SNR. We have trained our network with a dataset containing five different particle classes: Spots, ring-shaped and Janus-type particles as well as rod-like and elliptical particles (Fig. \ref{Fig:Figure4}, Dataset 5). A training set of 100000 images and a validation set of 20000 images was used where each image contains a randomized number of particles with randomized classes, positions, sizes/orientations and intensities. Details of the dataset as well as a set of sample images are available in {Supplement 1}, Section 3E. Fig. \ref{Fig:Figure4} illustrates the performance of the model. Different particle classes are correctly identified despite their different sizes, orientations and intensities. Even rod-like particles are properly distinguished from elliptical particles. The latter two particles are very difficult to distinguish in algorithmic approaches  \cite{Crocker1996, Shen2000, Cheezum2001, Rogers2007, Anthony2008, Parthasarathy2012, Chenouard2014, Wang2016, Yucel2017, Kapoor2019}. The proposed neural network based detection technique will, therefore, be advantageous, especially in cases with multiple species and heterogeneous optical contrast. 

\paragraph*{Extension to Additional Parameters  -- Orientation Detection}

While the above evaluation only refers to particle positions and particle classes, one may extend the network also to include other parameters. The orientation of objects becomes in particular interesting when particles lack spherical symmetry or have anisotropic optical properties. As can be seen for the elliptical and rod-like particles in Fig. \ref{Fig:Figure4} the orientation of objects with a 180$^{\circ}$ rotational invariance can be partly retrieved from the aspect ratio of the detected bounding boxes. Nevertheless, this yields an ambiguity of 90$^{\circ}$ since one cannot distinguish between, e.g., -45$^{\circ}$ and 45$^{\circ}$. In case of objects with no rotational invariance and a quadratic bounding box an orientation detection via the aspect ratio of bounding boxes is not possible at all.

\begin{figure}[!ht]
\centering
\includegraphics[width=0.475\textwidth]{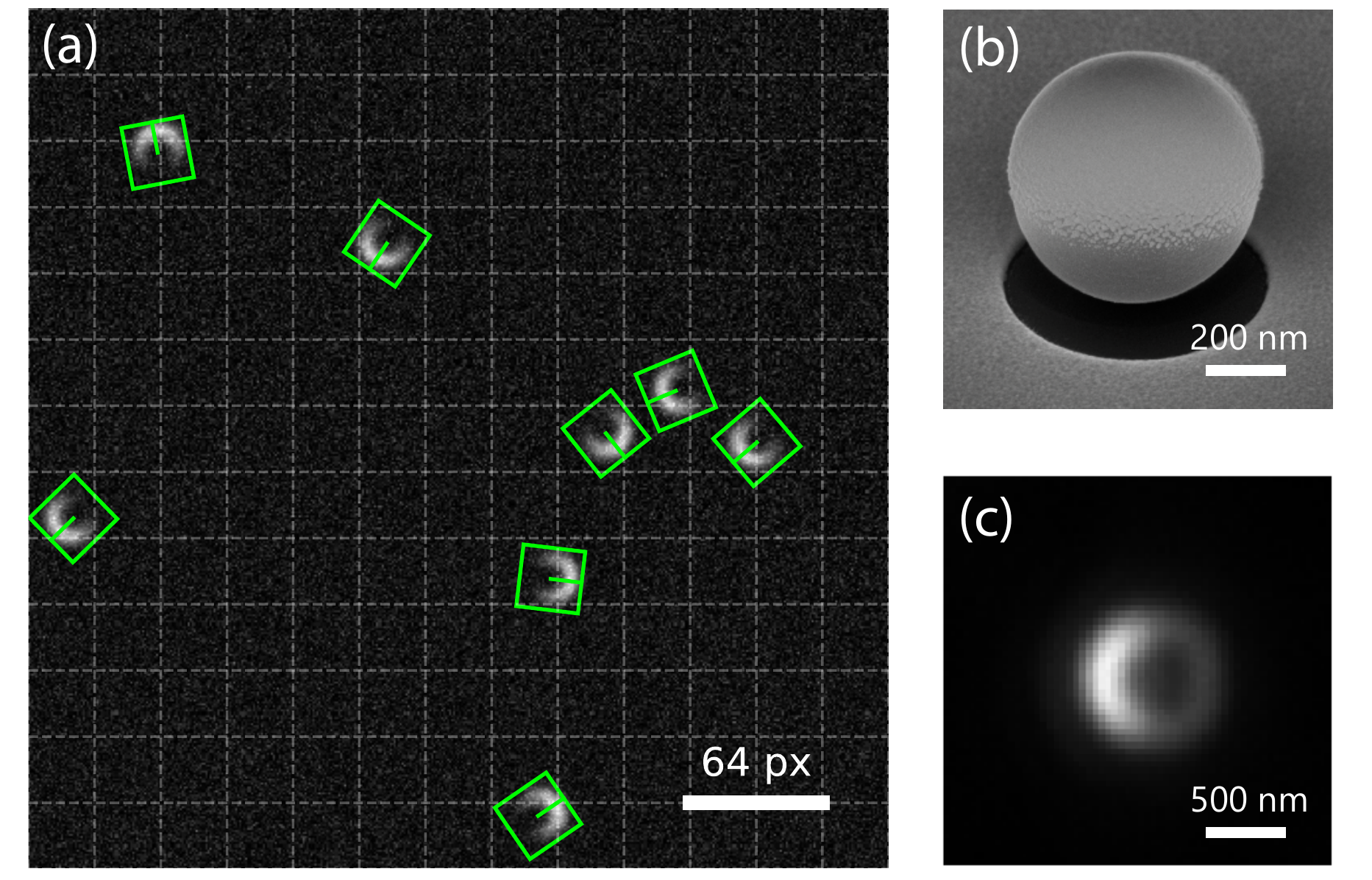}
\caption{Extention of YOLOv2 for orientation detection (YOLOv2.1). (a) Illustration of the predicted locations and orientations of synthetic Janus type particles. (b) A scanning electron microscopy (SEM) image of a 0.5~$\mathrm{\upmu m}$ diameter Janus particle. (c) Darkfield image of a 0.5~$\mathrm{\upmu m}$ diameter Janus particle.}
\label{Fig:Figure6}
\end{figure}

This is the case for Janus particles. Janus particles, when consisting of a hemispherical gold layer on top of a spherical polymer particle (Fig. \ref{Fig:Figure5}(b)) result in moon-shaped darkfield images (Fig. \ref{Fig:Figure6}(c)). The image, therefore, allows for a detection of the particle orientation but not from the bounding box.

To allow also for the prediction of the orientation the network needs to be modified and a new parameter, e.g., an angle $\varphi$, needs to be introduced to the loss function and the annotation format ({Supplement 1}, Section 9 and \cite{GitHub2020b}). Fig. \ref{Fig:Figure6}(b) shows the detection output of the extended network trained for Janus type particles \cite{Anthony2008, Qian2013, Bregulla2014}. An extension of the network to detect even more parameters such as the $z$-position of the particle or the out-of-plane rotation are easily possible again highlighting the flexibility of the network architecture \cite{Xiao2010, Anthony2006}. A more detailed description of the extended network, termed YOLOv2.1, can be found {Supplement 1},  Section 9.

\subsection{Experimental Real-Time Detection for Feedback Digital Microscopy}

\begin{figure*}[!ht]
\centering
\includegraphics[width=\textwidth]{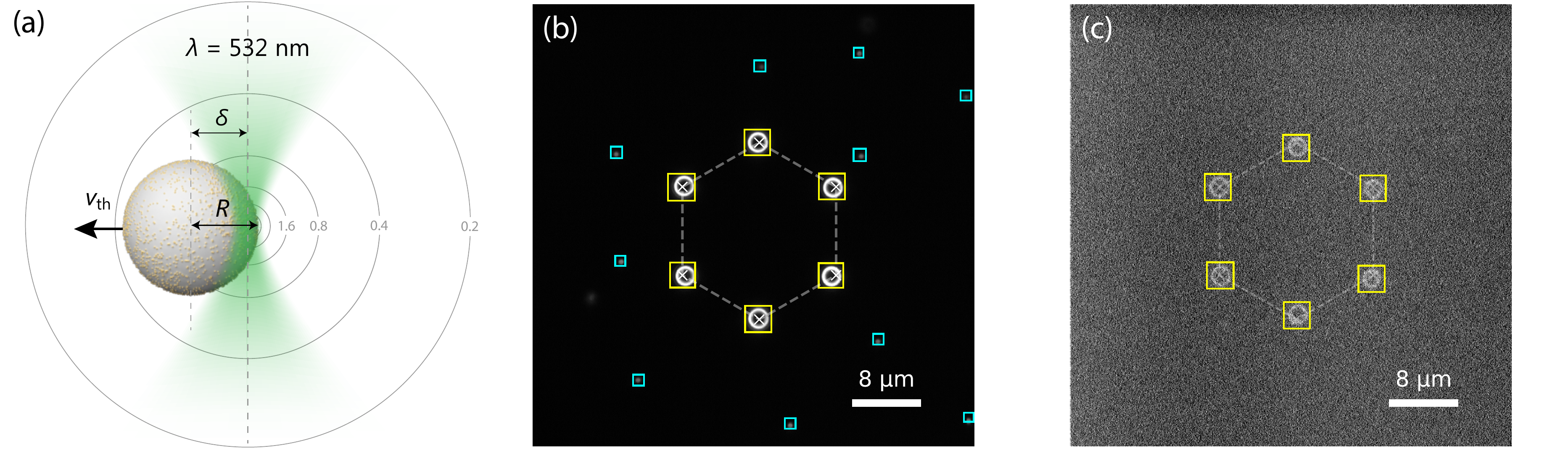}
\caption{Experimental real-time detection for feedback digital microscopy. (a) Sketch of a self-thermophoretic active particle composed of a melamine formaldehyde (MF) particle ($R = 1.1~\mathrm{\upmu m}$) covered with 10~nm gold nanoparticles with a surface coverage of about 30~\%. When asymmetrically heated with a focused laser ($\lambda = 532$~nm) an inhomogeneous surface temperature is generated resulting in a self-thermophoretic motion away from the laser focus. The velocity of the particle $v_\mathrm{th}$ depends on the incident laser power and on the displacement of the laser focus $\delta$ from the particle center. The highest velocity is observed for $\delta \approx R$. (b) Experimental feedback control of six individual active particles in a hexagonal pattern of six target positions ({Video 1}) at high SNR. (c) The same as in (b) at low SNR ({Video 2}).}
\label{Fig:Figure5}
\end{figure*}

Considering the inference time of 10 ms, the accuracy with an RMSE<1~px and the independence of the processing speed on the number of particles in the image, the above presented network is well suitable for real-time detection and feedback control of active particles.  For an experimental demonstration we studied the feedback controlled actuation of microparticles confined in a thin liquid film. The particle suspension contains 2.2~$\mathrm{\upmu m}$ diameter melamine formaldehyde (MF) particles as well as 0.5~$\mathrm{\upmu m}$ diameter polystyrene (PS) particles (Fig. \ref{Fig:Figure5}(b)).The surface of the MF particles is uniformly covered with gold nanoparticles of about 10~nm diameter with a surface coverage of about 30~\% (Fig. \ref{Fig:Figure5}(a)). The particles are observed using darkfield illumination with an oil-immersion darkfield condenser and a 100$\times$ oil-immersion objective. Details of the experimental setup and the sample preparation are available in {Supplement 1}, Sections 7, 8. The MF particles appear with a ring-shaped intensity profile whereas as the smaller PS particles have an approximate Gaussian intensity profile.

When illuminating the gold nanoparticles at the MF particle surface asymmetrically with a highly focused laser beam with a wavelength close to their plasmon resonance ($\lambda=532\, {\rm nm}$) an inhomogeneous surface temperature is generated. This inhomogeneous surface temperature is resulting in a self-thermophoretic propulsion away from the laser focus (Fig. \ref{Fig:Figure5}(a)) \cite{ Bregulla2014, Khadka2018}. 

To control the active particle motion direction the laser focus needs to be placed at the circumference of the particle in real time requiring the detection of the particles center position as fast as possible. This actuation scheme, which is similar to the photon nudging of Janus-type particles \cite{Qian2013, Bregulla2014} is achieved with the help of our neural network. 
For the experimental detection the network was trained with a two class dataset: Gaussian spots, as observed for the 0.5~$\mathrm{\upmu m}$ PS particles and ring-shaped intensity profiles as observed for the 2.2~$\mathrm{\upmu m}$ particles. To adapt for different magnifications the network was trained for different scales, e.g, different sizes of the two classes. The details of the dataset as well as a set of sample images are provided in {Supplement 1}, Section 3D. The laser in our setup is steered by an acousto-optic deflector. Multiple particles are addressed by quickly multiplexing the laser focus between different particle positions within the time of the inverse frame rate (20~ms). The camera was set to acquire images with a size of 512 $\times$ 512~pixels at an inverse frame rate of 20~ms (50~fps). To fit the network input the monochrome images from the camera are rescaled to 416 $\times$ 416~pixels and converted to grayscale RGB images. 

Fig. \ref{Fig:Figure5}(a) and {Video 1} demonstrate the control of six individual active particles (yellow boxes) with a background of passive gold nanoparticles (cyan boxes). We define six spatially fixed target positions arranged in a hexagon (white crosses in Fig. \ref{Fig:Figure6}(b) The particles are propelled by the feedback controlled laser towards the corresponding nearest point. As the resulting arrangement is constantly actuated the structure is dynamic as can be seen from {Video 1}. The structure can be maintained even at very low SNR when the gold nanoparticles are already undetectable (Fig. \ref{Fig:Figure5}(c), {Video 2}). 

\section{Summary}

In summary, we have shown that the adaption of a single-shot convolutional neural network (YOLOv2) is able to localize and classify several hundred objects in optical microscopy with an inference time of about 10~ms. The network training is implemented in Python/Keras using the TensorFlow backend. The real-time detection is achieved using a GPU supported TensorFlow C library that is usable in a programming language such as C++, LabVIEW or MATLAB. 
The speed of the classification and localization is independent of the number of particles. We analyzed the network performance with synthetic images and demonstrated its experimental application to feedback actuated self-thermophoretic active particles. Using this approach we envision further applications in the control of active matter and a combination with other machine learning techniques, e.g., reinforcement learning for an adaptive control and particle navigation \cite{Landin2018}. Also the control of processes in biological species is possible. The source code and scripts of our framework are open source and can be easily adapted and extended for such applications. 

\bibliography{References.bib} 

\end{document}